\documentstyle[epsfig,bibnorm]{lamuphys}
\begin{document}


\title{An Overview of the Sources for Electroweak Baryogenesis$^*$
}

\author{Tomislav\,Prokopec\inst{}
}

\institute{
Institut f\"ur Theoretische Physik, Universit\"at Heidelberg \\
          Philosophenweg 16, D-69120 Heidelberg, Germany
}

\maketitle

\begin{abstract}

After a short review of electroweak scale baryogenesis, 
we consider the dynamics of chiral fermions coupled to a complex
scalar field through the standard Yukawa interaction term
at a strongly first order electroweak phase transition.
By performing a systematic gradient expansion
we can use this simple model to study electroweak scale baryogenesis.
We show that the dominant sources for electroweak baryogenesis appear
at linear order in the Planck constant $\hbar$. We provide explicit
expressions for the sources both in the flow term and in the collision
term of the relevant kinetic Boltzmann equation. Finally, we indicate
how the kinetic equation sources appear in the fluid transport equations
used for baryogenesis calculations. 
 
\end{abstract}

\noindent
{\small $^*$
Presented at the `8th Adriatic meeting on particle physics in the new
millennium' (Dubrovnik, Croatia, 4 - 14 September 2001) 
          [Preprint: HD-THEP-02-18]
}

\section{Introduction}

The necessary requirements on dynamical baryogenesis at an epoch of the early
Universe are provided by the following Sakharov conditions:
\begin{itemize}
\item baryon number (B) violation
\item {\it charge} (C) and {\it charge-parity} (CP) violation
\item departure from thermal and kinetic equilibrium
\end{itemize}
The Sakharov conditions may be realised at the electroweak 
transition~\cite{KuzminRubakovShaposhnikov:1985}, 
provided the transition is strongly first order. Namely, 
C and CP violation are realised in the standard model (SM) for example 
through the Cabibbo-Kobayashi-Maskawa (CKM) matrix of quarks. B violation
is mediated through the Adler-Bell-Jackiw (ABJ) anomaly. At high temperatures
the ABJ anomaly is manifest via the unsuppressed sphaleron transitions,
and may be responsible for the observed baryon asymmetry today,
which is usually expressed as the baryon-to-entropy ratio:
\begin{equation}
 \frac{n_B}{s} = 3 - 7 \times 10^{-11}.
\label{observedB}
\end{equation}
This is obtained both as a nucleosynthesis constraint and from 
recent cosmic microwave background observations. 
 
 The standard model (SM) of elementary particles and interactions cannot
alone be responsible for the observed matter-antimatter 
asymmetry~(\ref{observedB}), primarily because the LEP bound on the Higgs 
mass $m_H\ga 112$~GeV is inconsistent with the requirement that the 
transition be strongly first order. A strongly first order transition is 
namely required in order for the baryons produced in the symmetric 
phase not be washed-out by the sphaleron transitions in the 
Higgs (`broken') phase. And this is so provided the transition is strong 
enough. This is usually expressed as the requirement $\Delta \phi\ga 1.1T$
on the jump in the Higgs expectation value $\phi$ of the phase transition
\cite{Shaposhnikov:1987}.

 Supersymmetric extensions of the Standard Model on the other hand
may result in a strongly first order transition. For example, in the Minimal
supersymmetric standard model the sphaleron bound can be satisfied
provided the stop and the lightest Higgs particles are not too heavy, 
$m_{\tilde t}\sim 120$~GeV and $m_{H}\la 120$~GeV~\cite{QuirosSeco:1999}.

 An efficient mechanism for baryon production at the electroweak phase 
transition is the 
{\it charge transport mechanism}~\cite{CohenKaplanNelson:1991}, which works
as follows. At a first order transition, when the Universe supercools,
the bubbles of the Higgs phase nucleate and grow. In presence of
a CP-violating condensate at the bubble interface, 
as a consequence of collisions of chiral fermions with 
scalar particles in presence of a scalar field condensate,
CP-violating currents are created
and transported into the symmetric phase, where they bias baryon number
production. The baryons thus produced are transported back into the Higgs
phase where they are frozen-in. The main unsolved problem of electroweak 
baryogenesis is systematic computation of the relevant CP-violating
currents generated at the bubble interface. Here we shall reformulate this 
problem in terms of calculating CP-violating sources in the kinetic
Boltzmann equations for fermions.

\begin{figure}
\centerline{
\epsfig{figure=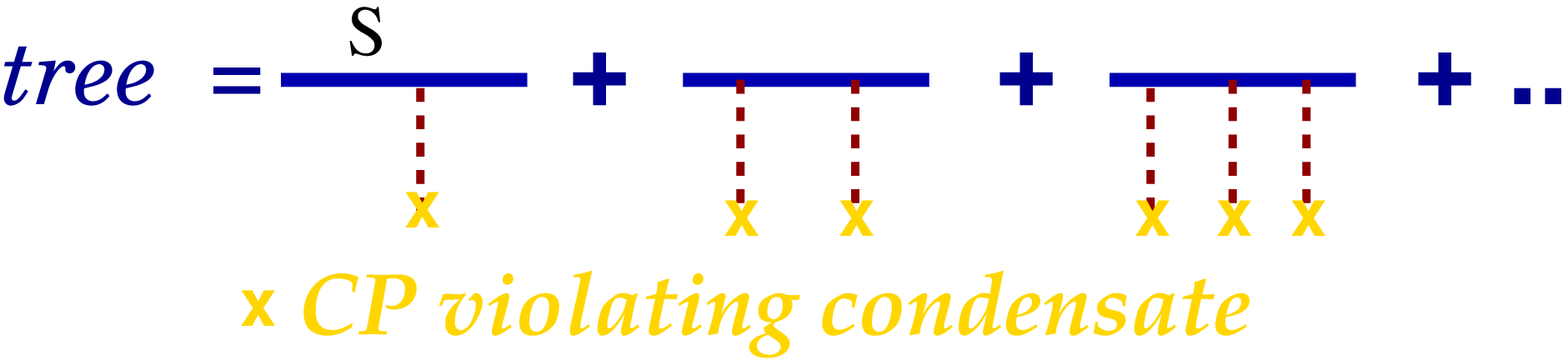, height=0.8in,width=3.5in}
}
\vskip -0.1in
\label{figure-I}
\caption{The tree level interactions of fermions with the scalar field 
condensate which, when expanded in gradients, lead to the CP-violating
semiclassical force in the kinetic Boltzmann equation.}
\end{figure}
\begin{figure}
\centerline{
\epsfig{figure=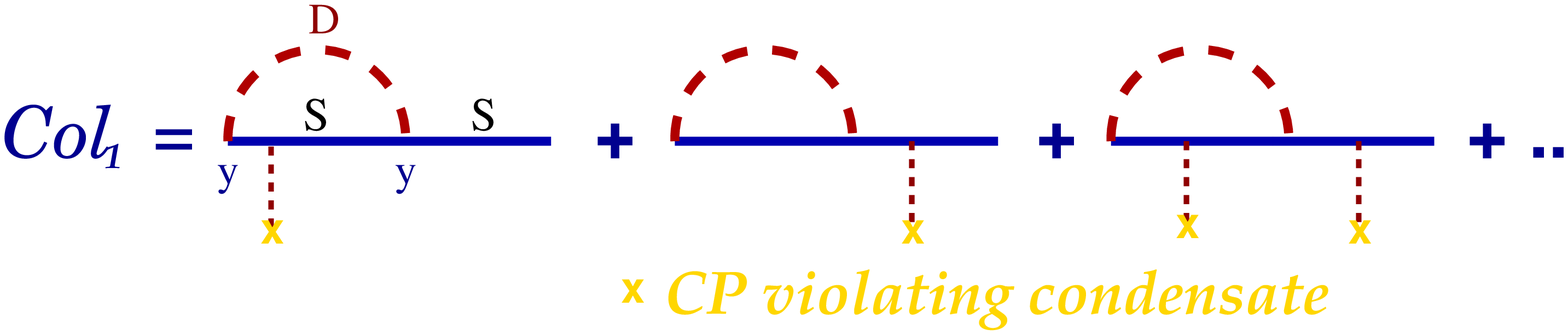, height=0.99in,width=4.5in}
}
\vskip -0.1in
\caption{The one-loop fermion-scalar diagrams, where interactions of fermions
with the scalar field condensate are explicitly shown. 
The condensate interactions, when expanded in gradients, result in
a CP-violating collisional source in the kinetic Boltzmann equation.
Upon projecting (or `cutting') the loop propagators on-shell one obtains
the CP-violating scalar particle absorption and emission processes.}
\label{figure-II}
\end{figure}

The techniques we report here are relevant for calculation of 
sources in the limit of thick phase boundaries and a weak coupling
to the Higgs condensate. In this case one can show that, to linear order
in the Planck constant $\hbar$, the quasiparticle picture for fermions
survives~\cite{KainulainenProkopecSchmidtWeinstock-I,
KainulainenProkopecSchmidtWeinstock-II}. In presence of a CP-violating 
condensate there are two types of sources: the semiclassical force 
in the flow term of the kinetic Boltzmann equation, and the collisional 
sources. The semiclassical force was originally introduced for baryogenesis
in two-Higgs doublet models in~\cite{JoyceProkopecTurok:1994},
and subsequently adapted to the chargino baryogenesis in the
Minimal Supersymmetric Standard Model (MSSM) 
in~\cite{ClineJoyceKainulainen:2000}. The semiclassical force corresponds
to tree level interactions with the condensate shown in figure~\ref{figure-I}
and it is universal in that its form is independent on interactions. 
The collisional sources on the other hand arise when fermions in the loop
diagrams interact with scalar condensates. In figure~\ref{figure-II} we
show typical CP-violating one-loop contributions to the collisional source. 
This source arises 
from one-loop diagrams in which fermions interact with a CP-violating 
scalar condensate. When viewed in the kinetic Boltzmann equation, 
these processes correspond to tree-level interactions in which fermions
absorb or emit scalar particles, whilst interacting in a CP-violating 
manner with the scalar condensate. The precise form of the collisional
source depends on the form of the interaction. In the following sections
we discuss how one can study the CP-violating collisional sources induced
by a typical Yukawa interaction term.

\section{Kinetic equations}

Here we work in the simple model of chiral fermions coupled to
a complex scalar field {\it via} the Yukawa interaction with the Lagrangian
of the form~\cite{KainulainenProkopecSchmidtWeinstock-I,
KainulainenProkopecSchmidtWeinstock-II}  
\begin{equation}
     {\cal L} = i\bar{\psi}{\mathbin{\partial\mkern-10.5mu\big/}}
\psi - \bar{\psi}_Lm\psi_R
           -  \bar{\psi}_Rm^*\psi_L  + {\cal L}_{\rm Yu},
\label{lagrangian0}
\end{equation}
where ${\cal L}_{\rm Yu}$ denotes the Yukawa interaction term
\begin{equation}
{\cal L}_{\rm Yu} = - y\phi \bar{\psi}_L\psi_R
           - y \phi^* \bar{\psi}_R\psi_L,
\label{lagrangian1}
\end{equation}
and $m$ is a complex, spatially varying mass term
\begin{equation}
     m(u) \equiv y' \Phi_0 = m_R(u) + i m_I(u) = |m(u)|\mbox{e}^{i\theta(u)}.
\label{mass1}
\end{equation}
Such a mass term arises naturally from an interaction with a scalar 
field condensate $\Phi_0 = \langle \hat \Phi(u)\rangle$. This 
situation is realised for example by the Higgs field condensate of a first
order electroweak phase transition in supersymmetric models.
When $\phi$ in~(\ref{lagrangian1}) is the Higgs field
the coupling constants $y$ and $y'$ coincide.
Our considerations are however not limited to this case.

\begin{figure}[htbp]
\leftline{\hspace{.3in} 
\epsfig{file=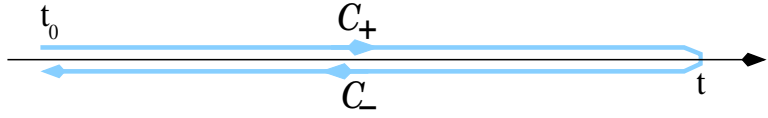, width=4.in,height=0.6in}
}
\vskip -0.1in
\caption{
The complex time contour for the Schwinger-Keldysh nonequilibrium formalism.
}
\label{figure-III}
\end{figure}

The dynamics of quantum fields can be studied by considering the equations 
of motion arising from the two-particle irreducible (2PI) effective action 
\cite{CornwallJackiwTomboulis:1974} in the Schwinger-Keldysh
closed-time-path formalism~\cite{Schwinger:1961,ChouSuHaoYu:1985}. 
This formalism is suitable for studying the dynamics of the non-equilibrium
fermionic and bosonic two-point functions 
\begin{eqnarray}
iS_{\alpha\beta}(u,v) 
 &=& \langle\Omega|T_{{\cal C}}
         \big[\psi_\alpha(u)\bar{\psi}_\beta(v)\big]|\Omega\rangle
\label{S}
\\
i\Delta(u,v) 
 &=& \langle\Omega|T_{{\cal C}}\big[\phi(u){\phi}^\dagger(v)\big]|\Omega\rangle
,
\label{Delta}
\end{eqnarray}
where $|\Omega\rangle$ is the physical state, and the time ordering
$T_{{\cal C}}$ is along the Schwinger contour shown in 
figure~\ref{figure-III}. For our purposes it suffices to consider the 
limit when $t_0\rightarrow -\infty$. The complex path time ordering 
can be conveniently represented in the Keldysh component formalism.
For example, for nonequilibrium dynamics of quantum fields 
the following Wightman propagator is relevant 
\begin{equation}
iS^<(u,v) 
 = - \langle\Omega|\bar{\psi}(v)\psi(u)|\Omega\rangle.
\label{Wightman}
\end{equation}

\begin{figure}[htbp]
\centerline{
\epsfig{figure=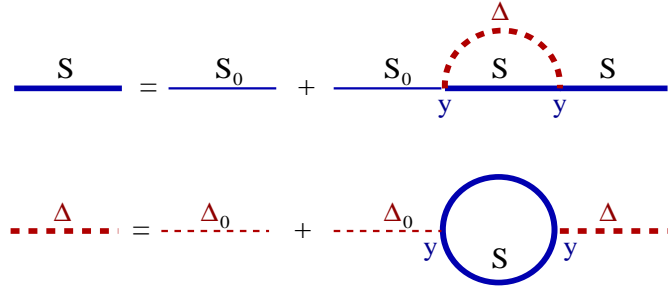, height=1.5in,width=3.5in}
}
\vskip -0.1in
\caption{The one-loop Schwinger-Dyson equations for the 
out-of-equilibrium fermionic ($S$) and scalar ($\Delta$) propagators. 
When projected on-shell and expanded in gradients, these equations reduce to
the kinetic Boltzmann equations.}
\label{figure-IV}
\end{figure}

For thick walls, that is for the plasma excitations whose de Broglie
wavelength $\ell_{\rm dB}$ is small in comparison to the phase interface
thickness $L_w$, it is suitable to work in the Wigner representation for
the propagators, which corresponds to the Fourier transform with respect
to the relative coordinate $r=u-v$, and expand in the gradients of
average coordinate $x=(u+v)/2$. This then represents an expansion in powers
of $\ell_{\rm dB}/L_w$. When written in this Wigner representation, 
the kinetic equations for fermions
become~\cite{KainulainenProkopecSchmidtWeinstock:2002c}
\begin{equation}
{\cal D}S^< \equiv 
   \Big(\frac i2\partial\!\!\!/+ k\!\!\! / 
 -(mP_R-m^*P_L) e^{{-\frac i2\stackrel{\leftarrow}{\partial}}\cdot\;\partial_k}
   \Big) S^{<}
    = {\cal C}_\psi,
\label{eom-S<}
\end{equation}
where for simplicity we neglected the contributions from self-energy 
corrections to the mass and the collisional 
broadening term~\cite{KainulainenProkopecSchmidtWeinstock:2002c}. 
When the collision term ${\cal C}_\psi$ is approximated at one loop,
equation~(\ref{eom-S<}) corresponds to the nonequilibrium 
fermionic Schwinger-Dyson equation shown in figure~\ref{figure-IV}. Since
the scalar equation (also shown in figure~\ref{figure-IV})
does not yield CP-violating sources at first order in
gradients~\cite{KainulainenProkopecSchmidtWeinstock-I, 
KainulainenProkopecSchmidtWeinstock:2002e}, we shall not discuss it here.

As the bubbles grow large, they tend to become more and more planar. Hence,
it suffices to consider the limit of a planar phase interface, in which 
the mass condensate in the wall frame becomes a function of
one coordinate only, $m=m(z)$. Further, we keep only 
the terms that contribute at order $\hbar$ to Eq.~(\ref{eom-S<}),
which implies that we need to keep second order gradients of the mass term
\begin{eqnarray}
m e^{{-\frac i2\stackrel{\leftarrow}{\partial}}\cdot\;\partial_k}
   = m + \frac i2 m' \partial_{k_z}  
    -  \frac 18 m'' \partial_{k_z}^2 + o(\partial_z^3), 
\label{m-hat}
\end{eqnarray}
where $m=m(z)$, $m' \equiv \partial_z m$ and $m'' \equiv \partial_z^2 m$. 
On the other hand, in the collision term  ${\cal C}_\psi$ we need to 
consider terms only up to linear order in derivatives 
\begin{eqnarray}
 {\cal C}_\psi  &=& {\cal C}_{\psi 0} + {\cal C}_{\psi 1} + ..
 \nonumber\\ 
  {\cal C}_{\psi 0} &=& - \frac 12
           \Big(\Sigma^>S^< - \Sigma^<S^> \Big)
 \nonumber\\ 
  {\cal C}_{\psi 1} &=& - \frac i4 
 (\partial_z^{(1)}\partial_{k_z}^{(2)} - \partial_{k_z}^{(1)}\partial_z^{(2)})
             \Big(\Sigma^>S^< - \Sigma^< S^>\Big),
\label{C-psi}
\end{eqnarray}
where $\Sigma^<$ and $\Sigma^>$ represent the fermionic self-energies, and 
the derivatives $\partial_{z}^{(1)}$, $\partial_{k_z}^{(1)}$ 
($\partial_z^{(2)}$, $\partial_{k_z}^{(2)}$) act on the first
(second) factor in the parentheses. 

 An important observation is that,
when $G = G(k_\mu,t-\vec x_\|\cdot\vec k_\|,z)$, the spin 
in the $z$-direction (of the interface motion)
\begin{equation}
  S_z  \equiv L^{-1}(\Lambda) \,\tilde{\!S}_z L(\Lambda)
  \; = \;  
 \gamma_\| \Big( \tilde S_z 
      - i(\vec v_\|\times \vec \alpha)_z \Big) 
\label{Sz}
\end{equation}
is conserved 
\begin{equation}
 [{\cal D},S_z]S^< = 0,
\label{spin-commutator}
\end{equation}
where ${\cal D}$ is the differential operator in Eq.~(\ref{eom-S<}),
$\vec\alpha = \gamma^0 \vec\gamma$, $\tilde S_z = \gamma^0\gamma^3\gamma^5$,
and $\gamma_\| = 1/(1- \vec v_\|^2)^{1/2}$.
This then implies that, without a loss of generality, the fermionic Wigner
function can be written in the following block-diagonal form 
\begin{eqnarray}
    S^< &=& \sum_{s=\pm} S_s^<
\nonumber\\
    S^< &=& L(\Lambda)^{-1}\tilde S^< L(\Lambda)
\nonumber\\
    - i\gamma^0\tilde{S}^{<}_s 
    &=& \frac{1}{4} 
   ({\mathbf{1}}+s\sigma^3) \otimes \rho^a\tilde{g}^s_a,
\label{Dec1+1}
\end{eqnarray}
where $\sigma^3$ and $\rho^i$ ($i=1,2,3$) are the Pauli matrices
and $\rho^0 = {\mathbf{1}}$ is the $2\times 2$ unity matrix
and $L(\Lambda)$ is the following Lorentz boost operator
\begin{equation}
  L(\Lambda) = \frac{k_0 + \tilde{k}_0 
               - \gamma^0\vec{\gamma}\cdot\vec{k}_{\|}}
                {\sqrt{2\tilde{k}_0(k_0+\tilde{k}_0)}},
\label{L}
\end{equation}
with $\tilde k_0 = {\rm sign}(k_0)({k_0^2 -\vec k_\|^2})^{1/2}$. 
The boost $\Lambda$ corresponds to a Lorentz transformation that 
transforms away $\vec k_\|$. 

 With the decomposition~(\ref{Dec1+1}) the trace of the antihermitean part
of Eq.~(\ref{eom-S<}) can be written as the following 
{\em algebraic} constraint 
equation~\cite{KainulainenProkopecSchmidtWeinstock-II} 
\begin{equation}
\left(k^2 - |m|^2 + \frac{s}{\tilde{k}_0}(|m|^2\theta') 
    \right) g^s_{00} = 0 ,
\label{constraint3+1}
\end{equation}
where $g^s_{00} = \gamma_\| \tilde g_0^{s<}$ denotes
the particle density on phase space $\{k_\mu,x_\nu\}$.
Equation~(\ref{constraint3+1}) has a spectral solution

\begin{eqnarray}
g^{s}_{00}  &\equiv& \sum_\pm \frac{2\pi }{Z_{s\pm}}
                 \, n_s \,\delta(k_0 \mp \omega_{s\pm}),
\label{spectral-dec}
\end{eqnarray}
where $\omega_{s\pm}$ denotes the dispersion relation
\begin{equation}
     \omega_{s\pm} = \omega_0 \mp s \frac{|m|^2\theta'}
   {2\omega_0\tilde\omega_0},
     \qquad \qquad 
\omega_0 = \sqrt{ \vec k^2 + |m|^2},
\qquad
\tilde\omega_0 = \sqrt{\omega_0^2-\vec k_{\parallel}^{\,2}}
\label{dispersion1}
\end{equation}
and $Z_{s\pm} = 
  1 \mp s|m|^2\theta'/2\tilde\omega_0^{3}$. The delta functions 
in~(\ref{spectral-dec}) project $n_s(k_\mu,t-\vec x_\|\cdot\vec k_\|,z)$ 
on-shell, thus yielding the distribution functions $f_{s+}$ and $f_{s-}$
for particles and antiparticles with spin $s$, respectively, defined by 
\begin{eqnarray}
     f_{s+} &\equiv& n_s(\omega_{s+},k_z,t-\vec x_\|\cdot\vec k_\|,z) \nonumber
 \\
     f_{s-} &\equiv& 1 - n_s(-\omega_{s-},-k_z,t+\vec x_\|\cdot\vec k_\|,z).
\label{fs}
\end{eqnarray}
This on-shell projection proves the implicit
assumption underlying the semiclassical WKB-methods, that the plasma
can be described as a collection of single-particle excitations with
a nontrivial space-dependent dispersion relation. 
In fact, the decomposition~(\ref{Dec1+1}), Eq.~(\ref{constraint3+1}) and 
the subsequent discussion imply that the physical states that correspond
to the quasiparticle plasma excitations are the eigenstates of the spin
operator~(\ref{Sz}). 

 Taking the trace of the Hermitean part of Eq.~(\ref{eom-S<}), integrating
over the positive and negative frequencies and taking account
of~(\ref{spectral-dec}) and~(\ref{fs}), one obtains 
the following on-shell kinetic 
equations~\cite{KainulainenProkopecSchmidtWeinstock-II}
\begin{equation}
        \partial_t f_{s\pm} + \vec v_\|\cdot \nabla_\| f_{s\pm}
      +  v_{s\pm} \partial_z f_{s\pm}
      + F_{s\pm} \partial_{k_z} f_{s\pm} = {\cal C}_{\psi s\pm}[f_{s\pm}],
\label{ke-fspm0}
\end{equation}
where $f_{s\pm} = f_{s\pm}(k_\mu, z, \, t - \vec v_\| \cdot \vec x_\|)$,
${\cal C}_{s\pm}[f_{s\pm}]$ is the collision term obtained by 
integrating~(\ref{C-psi}) over the positive and negative frequencies, 
respectively, the quasiparticle {\it group velocity} 
$v_{s\pm} \equiv k_z/\omega_{s\pm}$ is expressed in terms of the kinetic
momentum $k_z$ and the quasiparticle energy 
$\omega_{s\pm}$~(\ref{dispersion1}), and the {\em semiclassical force} 
\begin{eqnarray}
    F_{s\pm} = - \frac{{|m|^2}^{\,\prime} }{2\omega_{s\pm}}
                   \pm  \frac{s(|m|^2\theta^{\,\prime})^{\,\prime}}
                   {2\omega_0\tilde\omega_0}.
\label{Fspm}
\end{eqnarray}
In the stationary limit in the wall frame the distribution function
simplifies to $f_{s\pm} = f_{s\pm}(k_\mu, z,)$.
When compared with the 1+1 dimensional case studied 
in~\cite{KainulainenProkopecSchmidtWeinstock-I} the sole, but significant,
difference in the force~(\ref{Fspm}) is that the CP-violating 
$\theta'$-term is enhanced by the boost-factor $\gamma_\parallel 
= \omega_0/\tilde\omega_0$,
$\tilde \omega_0 = ({\omega_0^2-\vec k_\parallel^{\,2}})^{1/2}$, which,
when integrated over the momenta, leads to an enhancement by about a 
factor {\it two} in the CP-violating source from the semiclassical force. 

\section{Sources for baryogenesis in the fluid equations}

\begin{figure}
\centerline{
\epsfig{figure=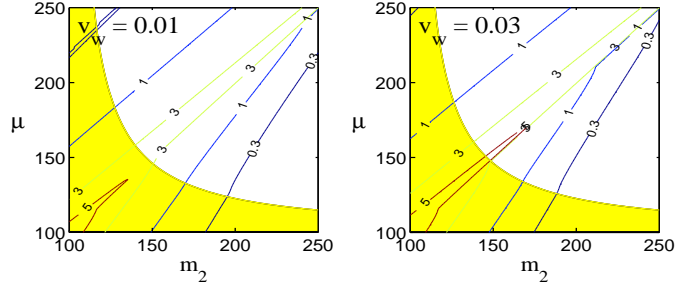, height=2.1in,width=3.5in}
}
\vskip -0.4in
\caption{The semiclassical force baryogenesis mediated by charginos
of the MSSM calculated in the helicity basis. 
The figures show contours for the baryon-to-entropy ratio in the units 
of $10^{-11}$ for two wall velocities $v_w = 0.01$ and
$v_w = 0.03$ as a function of the soft susy breaking parameters $\mu$ and 
$m_2$. A maximal CP violation in the chargino sector is assumed.
The shaded (yellow) regions are ruled out by the LEP measurements. 
The observed baryon asymmetry is in these units $3-7$. 
(The figure is taken from the latter reference in~[8].
)}
\label{figure-VI}
\end{figure}

 Fluid transport equations are usually obtained by taking first two
moments of the Boltzmann transport equation~(\ref{ke-fspm0}). That is, 
integrating Eq.~(\ref{ke-fspm0}) over the spatial momenta results in 
the continuity equation for the vector current, while multiplying by 
the velocity and integrating over the momenta yields the Euler equation.
The physical content of these equations can be summarized as the particle
number and fluid momentum density conservation laws for fluids, respectively. 
This procedure is necessarily approximate simply because the fluid equations
describe only very roughly the rich momentum dependence described by the 
distribution functions of the Boltzmann equation~(\ref{Fspm}). 
The fluid equations can be easily reduced to the diffusion equation
which has so far being used almost exclusively for electroweak baryogenesis
calculations at a first order electroweak phase transition. A useful 
intermediate step in derivation of the fluid equations is
rewriting Eq.~(\ref{ke-fspm0}) for the CP-violating departure from equilibrium
$\delta f_{si} = \delta f_{si+} - \delta f_{si-}$ as follows
\begin{eqnarray}
  &&  \left(\partial_t  + \frac{k_z}{\omega_{0i}} \partial_z 
  - \frac{{|m_i|^2}'}{2\omega_{0i}} \partial_{k_z}\right) \delta f_{si} 
  + v_w \delta F_{si} (\partial_{\omega} f_{\omega})_{\omega_{0i}} 
\nonumber\\
  &+& v_w F_{0i}\delta\omega_{si}
 \bigg[\Big(\frac{\partial_{\omega} f_{\omega}}{\omega}\Big)_{\omega_{0i}}
      - (\partial^2_{\omega} f_{\omega})_{\omega_{0i}} \bigg]
  = {\cal C}_{\psi si},
\label{boltzeqn-split-2}
\end{eqnarray}
where $i$ is the species (flavour) index, 
$f_\omega = 1/(e^{\beta\omega}+1)$, and 
%
\begin{eqnarray}
    F_{0i} &=& - \frac{{|m_i|^2}'}{2\omega_{0i}}
\nonumber\\
 \delta\omega_{si} &=& s\frac{(|m_i|^2\theta_i)'}{\omega_{0i}\tilde\omega_{0i}}
\nonumber\\
\delta F_{si} &\equiv& F_{si+} - F_{si-} 
              = s\frac{(|m_i|^2\theta_i')'}{\omega_{0i}\tilde\omega_{0i}}
\nonumber\\
   {\cal C}_{\psi si} &\equiv& {\cal C}_{\psi si+} - {\cal C}_{\psi si-}.
\label{boltzeqn-split-3}
\end{eqnarray}
When integrating~(\ref{boltzeqn-split-2}) over the momenta, 
the flow term yields two sources in the continuity equation for the vector
current. The former comes from the CP-violating spin dependent semiclassical
force, and has the form
\begin{eqnarray}
 {\cal S}^a_{si} &=& v_w \int \frac{d^3k}{(2\pi)^3}\delta F_{si} 
             (\partial_{\omega} f_{\omega})_{\omega=\omega_{0i}}
\nonumber\\
               &=& -sv_w \frac{(|m_i|^2\theta'_i)'}{4\pi^2}\; {\cal J}_a(x_i)
\label{boltzeqn-split-6a}
\end{eqnarray}
with $x_i=|m_i|/T$, while the latter comes from the CP-violating shift
in the quasiparticle energy, and can be written as 
\begin{eqnarray}
 {\cal S}^b_{si} &=& v_w \int \frac{d^3k}{(2\pi)^3}
F_{0i}\delta\omega_{si}
 \Big[\big(\partial_{\omega} f_{\omega}/\omega\big)_{\omega_{0i}}
      - (\partial^2_{\omega} f_{\omega})_{\omega_{0i}} \Big]
\nonumber\\
        &=& sv_w \frac{(|m_i|^2\theta'_i)}{2\pi^2T}{|m_i|}'\;{\cal J}_b(x_i).
\label{boltzeqn-split-6b}
\end{eqnarray}
\begin{figure}[htbp]
\begin{center}
\epsfig{file=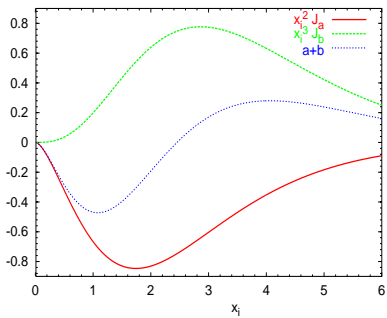, height=1.7in,width=2.2in}
\end{center}
\vskip -0.26in
\caption[fig1]{%
\small
The flow term sources~(\ref{boltzeqn-split-6a})-(\ref{boltzeqn-split-6b}) 
characterised by the integrals $x_i^2{\cal J}_a(x_i)$ ({\it red solid}) and
$x_i^3{\cal J}_b(x_i)$ ({\it green dashed}) as a function
of the rescaled mass $x_i=|m_i|/T$. The sum of the two sources
({\it thin blue line}) is also shown.
}
\label{figure-V}
\end{figure}
The total source is simply the sum of the two,
${\cal S}_{si} = {\cal S}^a_{si} + {\cal S}^b_{si}$.
To get a more quantitative understanding of these sources, in
figure~\ref{figure-V} we plot the integrals ${\cal J}_a$ and ${\cal J}_b$
in equations~(\ref{boltzeqn-split-6a}) and~(\ref{boltzeqn-split-6b}).
A closer inspection of the sources ${\cal S}_{si}^a$ and ${\cal S}_{si}^b$
indicates that the total source ${\cal S}_{si}$ can be also rewritten as
the sum of two sources: the source $\propto {|m_i|^2}'\Theta_i'$, 
characterized by $x_i^2{\cal J}_a + x_i^3 {\cal J}_b$, and the source
$\propto {|m_i|^2}\Theta_i''$, characterized by $x_i^2{\cal J}_a$.
We note that in the spin state quasiparticle basis the flow term
sources appear in the continuity equation for the vector current,
while in the helicity basis, which is usually used in 
literature~\cite{JoyceProkopecTurok:1994,ClineJoyceKainulainen:2000}, 
the flow term sources appear in the Euler equation. 
In figure~\ref{figure-VI} we show recent results of baryogenesis 
calculations of Ref.~\cite{ClineJoyceKainulainen:2000} based on the 
CP-violating contribution to the semiclassical force in the chargino sector
of the Minimal Supersymmetric Standard Model (MSSM). This calculation is 
based on the quasiparticle picture based on helicity states.
The analysis suggests that one can dynamically obtain
baryon production marginally consistent with the observed
value~(\ref{observedB}), provided $m_2\sim \mu \sim 150$~GeV
and $v_w \sim 0.03 c$.

\begin{figure}
\centerline{
\epsfig{figure=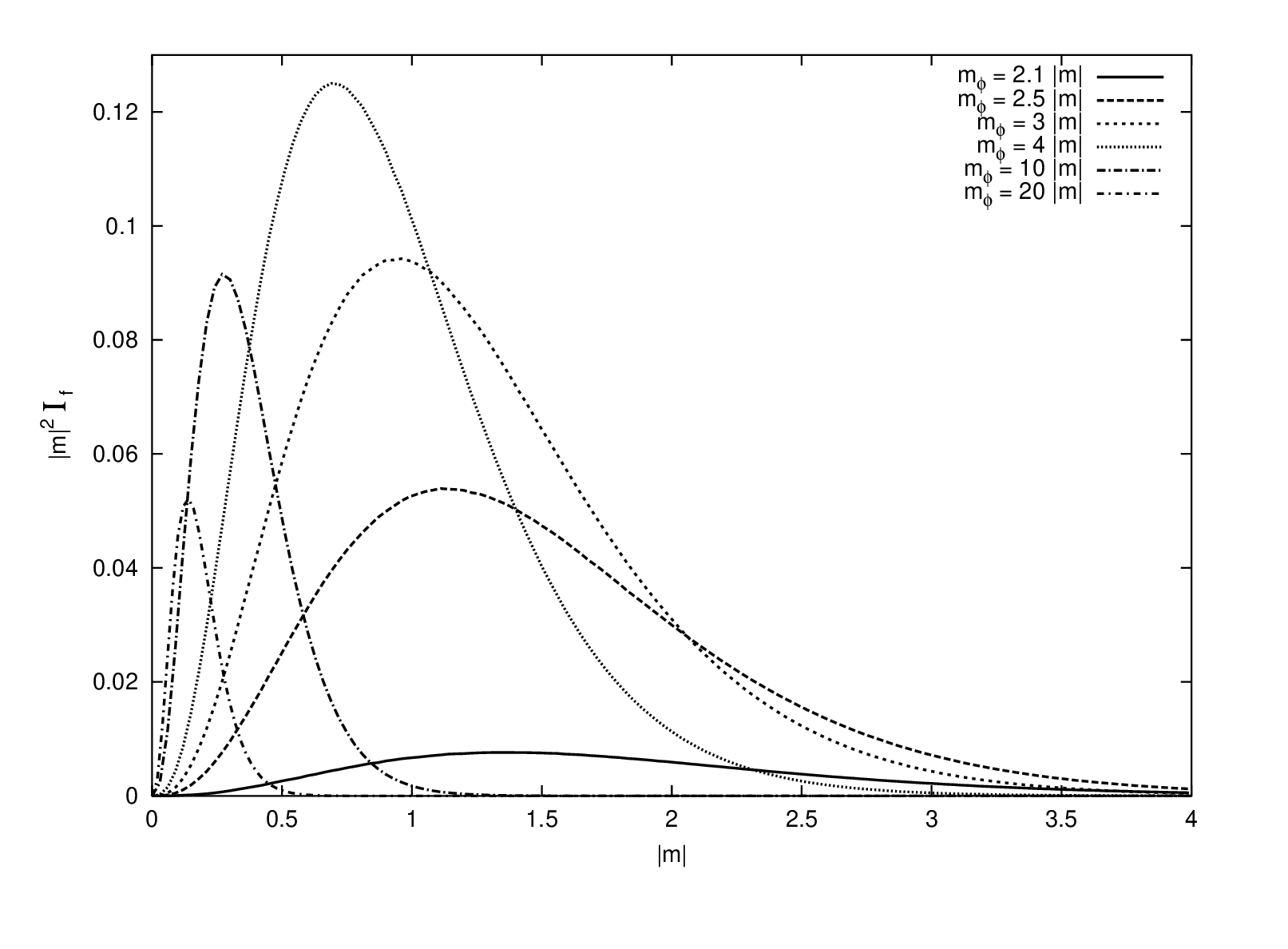, height=2.2in,width=3.in}
}
\vskip -0.3in
\caption{ 
The collisional source contributing to the fermionic 
kinetic equation at one loop for the mass ratios
$m_\phi/|m| = 2.1, 2.5, 3, 4, 10$ and 20, respectively. 
The source peaks for $|m| \approx 0.7T$ and $m_\phi\approx 4|m|$.
}
\label{figure-VII}
\end{figure}

 We now turn to discussion of the collision term sources 
in Eqs.~(\ref{ke-fspm0}) and~(\ref{boltzeqn-split-2}). 
We assume that the self-energies $\Sigma^{>,<}$ are approximated 
by the one-loop expressions ({\it cf.} figure~\ref{figure-IV})
\begin{eqnarray}
 \Sigma^{<,>}(k,x) &=& i y^2\int \frac{d^4k'd^4k''}{(2\pi)^8}\,\big[
(2\pi)^4\delta(k-k'+k'') P_L S^{<,>}(k',x) P_R \Delta^{>,<}(k'',x) 
\nonumber\\
&& \qquad\qquad
  + (2\pi)^4\delta(k-k'-k'') P_R S^{<,>}(k',x) P_L \Delta^{<,>}(k'',x)
\big],
\label{Sigma-<>}
\end{eqnarray}
where $\Delta^{<}$ and $\Delta^{>}$ denote the bosonic Wigner functions.
This expression contains both the CP-violating sources and 
relaxation towards equilibrium. The CP-violating sources can be 
evaluated by approximating the Wigner functions 
$S^{>,<}$ and $\Delta^{>,<}$ by the equilibrium expressions accurate to 
first order in derivatives. The result of the investigation is as follows.
There is no source contributing to the continuity equation, while the source
arising in the Euler equation is of the 
form~\cite{KainulainenProkopecSchmidtWeinstock:2002c}
\begin{eqnarray}
2\int_{\pm}\frac{d^4k}{(2\pi)^4}\frac{k_z}{\omega_0} {\cal C}_{\psi si} 
 &=& v_w y^2\frac { s|m|^2\theta'}{32\pi^3 T}
{\cal I}_f(|m|,m_\phi),
\label{FermionicSource1}
\end{eqnarray}
where the function ${\cal I}_f(|m|,m_\phi)$ is plotted in 
figure~\ref{figure-VII}. It is encouraging that the source vanishes 
for small values of the mass parameters, which suggests that the expansion
in gradients we used here may yield the dominant sources. Note that the 
source is nonvanishing only in the kinematically allowed region, 
$m_\phi\geq 2 |m|$. When the masses are large, $|m|, m_\phi \gg T$, 
the source is as expected Boltzmann suppressed. 
It would be of interest to make a detailed comparison between the sources
in the flow term and those in the collision term. This is a subject of 
an upcoming publication.
%

%
%

\end{document}